\documentclass[twocolumn,preprintnumbers,pra,hyperref]{revtex4}
\usepackage{amssymb}
\usepackage{amsfonts}
\usepackage{amsmath}
\usepackage{graphicx}

\begin{document}

\title{Fresnel operator, squeezed state and Wigner function for
Caldirola-Kanai Hamiltonian}
\author{Shuai Wang$^{\ast }$, Hong-Yi Fan, Hong-Chun Yuan$^{\dagger }$}
\affiliation{{\small Department of Physics, Shanghai Jiao Tong University, Shanghai
200030, China. }\\
$^{\ast }${\small wangshuai197903@sohu.com} \\
$^{\dagger }${\small Corresponding author: yuanhch@126.com or
yuanhch@sjtu.edu.cn}}

\begin{abstract}
Based on the technique of integration within an ordered product (IWOP) of
operators we introduce the Fresnel operator for converting Caldirola-Kanai
Hamiltonian into time-independent harmonic oscillator Hamiltonian. The
Fresnel operator with the parameters $A,B,C,D$ corresponds to classical
optical Fresnel transformation, these parameters are the solution to a set
of partial differential equations set up in the above mentioned converting
process. In this way the exact wavefunction solution of the Schr\"{o}dinger
equation governed by the Caldirola-Kanai Hamiltonian is obtained, which
represents a squeezed number state. The corresponding Wigner function is
derived by virtue of the Weyl ordered form of the Wigner operator and the
order-invariance of Weyl ordered operators under similar transformations.
The method used here can be suitable for solving Schr\"{o}dinger equation of
other time-dependent oscillators.


\end{abstract}

\maketitle

\section{Introduction}

Damped harmonic oscillator is a typical example of dissipative systems.
Usually people introduce a time-dependent Hamiltonian for describing such a
dissipative system. The Caldirola-Kanai (CK) Hamiltonian \cite{01,02} model
for the damped harmonic oscillator has brought considerable attention in the
past few decades because it offers many applications in various areas of
physics. In order to obtain the exact solution to the Schr\"{o}dinger
equation for CK Hamiltonian, several techniques, such as path integral and
propagator method, dynamical invariant operator method, etc, are used \cite%
{03,04,05,06,07,08,09,10}. In this work following Dirac's idea "... for a
quantum dynamic system that has a classical analogue, unitary transformation
in the quantum theory is the analogue of the contact transformation in the
classical theory..." , we shall adopt a new approach for treating CK
Hamiltonian, i.e., to construct a so-called Fresnel operator to convert the
Hamiltonian of explicitly time-dependent oscillator into time-independent
harmonic oscillator Hamiltonian. The Fresnel operator with parameters $%
A,B,C,D$ corresponds to a classical optical Fresnel transformation, and
these parameters are the solution to a set of partial differential equations
set up in the above mentioned converting process. In this way the exact
time-dependent wavefunction of the Schr\"{o}dinger equation governed by the
CK Hamiltonian can be directly obtained, which represents a squeezed number
state. The corresponding Wigner function is derived by virtue of the Weyl
ordered form of the Wigner operator \cite{11} and the order-invariance of
Weyl ordered operators under similar transformations \cite{12}.

\section{Fresnel operator as mapping of classical canonical transformation
in coherent state representation}

By mapping $\left( x,p\right) \rightarrow \left(
A(t)x+B(t)p,C(t)x+D(t)p\right) ,$ where $AD-BC=1$ is kept in time evolution,
in the canonical coherent state representation \cite{13}%
\begin{equation}
\left \vert \left(
\begin{array}{c}
x \\
p%
\end{array}
\right) \right \rangle =\exp \left[ -\frac{1}{4}(x^{2}+p^{2})+\frac{1}{\sqrt{%
2}}(x+\text{i}p)a^{\dagger}\right] \left \vert 0\right \rangle ,  \label{1}
\end{equation}
where $a^{\dagger}$ is the bosonic creation operator with $\left[
a,a^{\dagger}\right] =1$, Fan et al\textit{\ }\cite{14,15} have set up an
explicit quantum mechanical unitary operator
\begin{align}
F & =\sqrt{\frac{1}{2}\left[ A+D-\text{i}(B-C)\right] }  \notag \\
& \times \int \frac{dxdp}{2\pi}\left \vert \left(
\begin{array}{cc}
A(t) & B(t) \\
C(t) & D(t)%
\end{array}
\right) \left(
\begin{array}{c}
x \\
p%
\end{array}
\right) \right \rangle \left \langle \left(
\begin{array}{c}
x \\
p%
\end{array}
\right) \right \vert .  \label{2}
\end{align}
Using the technique of integration within an ordered product (IWOP) of
operators \cite{14,15,16} to perform the integration in Eq.(\ref{2}) yields
\begin{align}
F & =\exp \left \{ {{{{{\frac{A-D+\text{i}(B+C)}{2[A+D+\text{i}(B-C)]}%
a^{\dagger2}}}}}}\right \}  \notag \\
& \times \exp \left[ (a^{\dagger}a+\frac{1}{2})\ln \left( \frac{2}{A+D+\text{%
i}(B-C)}\right) \right]  \notag \\
& \times \exp \left \{ {{{{{-\frac{A-D-\text{i}(B+C)}{2[A+D+\text{i}(B-C)]}%
a^{2}}}}}}\right \} .  \label{3}
\end{align}
Taking the matrix element of operator $F$ in the coordinate representation $%
\left \vert x\right \rangle $ , the result is given by \cite{16}%
\begin{equation}
\left \langle x^{\prime}\right \vert F\left \vert x\right \rangle =\frac{1}{%
\sqrt{2\pi \text{i}B}}\exp \left[ \frac{\text{i}}{2B}\left(
Ax^{2}-2x^{\prime}x+Dx^{\prime2}\right) \right] ,  \label{4}
\end{equation}
which is just the integration kernel of optical Fresnel transforation \cite%
{17}, thus $F$ is named as Fresnel operator, i.e. to correspond to a Fresnel
transformation in Fourier optics. Further, it has been proved that $F$ has
its canonical operator $(X,P)$ representation \cite{18}%
\begin{align}
F & =\exp \left( \frac{\text{i}C}{2A}X^{2}\right) \exp \left[ -\frac {\text{i%
}}{2}(XP+PX)\ln A\right]  \notag \\
& \times \exp \left( -\frac{\text{i}B}{2A}P^{2}\right) ,  \label{5}
\end{align}
where $X=\frac{a+a^{\dagger}}{\sqrt{2}}$ and $P=\frac{a-a^{\dagger}}{\text{i}%
\sqrt{2}}.$ One can see that $F$ is a general SU(1,1) single-mode squeezing
operator, because $X^{2}$, $P^{2}$ and $XP+PX$ are three generators of
SU(1,1) Lie algebra. The following relations can be easily obtained
\begin{align}
FXF^{\dagger} & =D(t)X-B(t)P,F^{\dagger}XF=A(t)X+B(t)P,  \label{6} \\
FPF^{\dagger} & =A(t)P-C(t)X,F^{\dagger}PF=C(t)X+D(t)P.  \notag
\end{align}

\section{Fresnel transformation for quantum mechanical time-dependent
oscillator}

In this section, we employ time-dependent Fresnel operator to study the
dynamic evolution of time-dependent harmonic oscillators. For a general
time-dependent Hamiltonian

\begin{equation}
\hat{H}=f(t)\frac{P^{2}}{2m}+g(t)\frac{1}{2}m\omega_{0}^{2}X^{2},  \label{7}
\end{equation}
with the Schr\"{o}dinger equation
\begin{equation}
\text{i}\frac{\partial \left \vert \psi(t)\right \rangle }{\partial t}%
=H\left \vert \psi(t)\right \rangle ,  \label{8}
\end{equation}
here $\hbar=1$, we hope that this time-dependent $H$ can be converted into a
time-independent harmonic oscillator by some time-dependent Fresnel
transformation, where $A(t),B(t),C(t),D(t)$ are determined by solving a
coupled partial differential equations, this can be derived as follows. By
performing a transformation on $\left \vert \psi(t)\right \rangle $ with
Fresnel operator, we have%
\begin{equation}
\left \vert \phi \right \rangle =F\left \vert \psi(t)\right \rangle ,
\label{9}
\end{equation}
which follows%
\begin{align}
\text{i}\frac{\partial \left \vert \phi \right \rangle }{\partial t} & =%
\text{i}\frac{\partial \left( F\left \vert \psi(t)\right \rangle \right) }{%
\partial t}  \notag \\
& =\text{i}\frac{\partial F}{\partial t}\left \vert \psi(t)\right \rangle
+FH\left \vert \psi(t)\right \rangle \equiv \mathcal{H}\left \vert \phi
\right \rangle ,  \label{10}
\end{align}
due to $FF^{-1}=1$ as well as $\frac{\partial F}{\partial t}F^{-1}+F^{-1}%
\frac{\partial F}{\partial t}=0,$ we have%
\begin{equation}
\mathcal{H=}FHF^{-1}-\text{i}\hbar F\frac{\partial F^{-1}}{\partial t}.
\label{11}
\end{equation}
In order to know $\mathcal{\hat{H}}$, we must calculate $\frac{\partial
F^{-1}}{\partial t}$, in fact, using Baker-Hausdorff formula
\begin{equation}
e^{\mathcal{A}}\mathcal{B}e^{-\mathcal{A}}=\mathcal{B+[A},\mathcal{B]+}\frac{%
1}{2!}[\mathcal{A},\mathcal{[A},\mathcal{B]}]+\cdots  \label{12}
\end{equation}
and Eq. (\ref{6}), we have
\begin{widetext}
\begin{align}
\frac{\partial F^{-1}}{\partial t} & =\frac{\partial}{\partial t}\left[ \exp(%
\frac{iB}{2A}P^{2})\exp[\frac{i}{2}(XP+PX)\ln A]\exp \left( -\frac {iC}{2A}%
X^{2}\right) \right]  \notag \\
& =F^{-1}\frac{i}{2}(\frac{1}{A}\frac{\partial B}{\partial t}-\frac{B}{A^{2}}%
\frac{\partial A}{\partial t})\left( AP-CX\right) ^{2}  \notag \\
& +F^{-1}\frac{i}{2}\frac{\partial A}{A\partial t}\newline
(XP+PX-\frac{2C}{A}X^{2})+F^{-1}\frac{-iX^{2}}{2}(\frac{1}{A}\frac{\partial C%
}{\partial t}-\frac{C}{A^{2}}\frac{\partial A}{\partial t})  \notag \\
& =\frac{i}{2}F^{-1}\left[ (A\frac{\partial B}{\partial t}-B\frac{\partial A%
}{\partial t})P^{2}+(C\frac{\partial D}{\partial t}-D\frac{\partial C}{%
\partial t})X^{2}+(D\frac{\partial A}{\partial t}-C\frac{\partial B}{%
\partial t})(XP+PX)\right] .   \label{13}
\end{align}
where we have considered $AD-BC=1$ as well as $A\frac{\partial D}{\partial t}%
+D\frac{\partial A}{\partial t}-B\frac{\partial C}{\partial t}-C\frac{%
\partial B}{\partial t}=0$. Substituting Eqs. (\ref{6}), (\ref{7}) and (\ref%
{13}) into Eq.(\ref{11}), we see%
\begin{align}
\mathcal{H}\text{ } & \mathcal{=}\frac{f(t)}{2m}FP^{2}F^{-1}+\frac {%
m\omega_{0}^{2}g(t)}{2}FX^{2}F^{-1}-\text{i}\hbar F\frac{\partial F^{-1}(t)}{%
\partial t}  \notag \\
& =\left[ \frac{A^{2}f(t)}{2m}\mathbf{+}\frac{m\omega_{0}^{2}g(t)}{2}B^{2}%
\mathbf{+}\frac{A\frac{\partial B}{\partial t}-B\frac{\partial A}{\partial t}%
}{2}\right] P^{2}  \notag \\
& +\left[ \frac{f(t)C^{2}}{2m}+\frac{m\omega_{0}^{2}g(t)}{2}D^{2}+\frac{C%
\frac{\partial D}{\partial t}-D\frac{\partial C}{\partial t}}{2}\right] X^{2}
\notag \\
& -\left[ \frac{ACf(t)}{2m}+\frac{m\omega_{0}^{2}g(t)DB}{2}-\frac {D\frac{%
\partial A}{\partial t}-C\frac{\partial B}{\partial t}}{2}\right] \left(
XP+PX\right) .   \label{14}
\end{align}
\end{widetext}
If we demand%
\begin{equation}
\mathcal{H=}\frac{P^{2}}{2m}+\frac{1}{2}m\omega^{2}X^{2},  \label{15}
\end{equation}
a time-independent Hamiltonian, where $\omega$ is to be determined shortly
later, we can derive the following coupled partial differential equations%
\begin{equation}
\frac{A^{2}f(t)}{2m}+\frac{B^{2}m\omega_{0}^{2}g(t)}{2}+\frac{1}{2}\left( A%
\frac{\partial B}{\partial t}-B\frac{\partial A}{\partial t}\right) =\frac{1%
}{2m},  \label{16}
\end{equation}%
\begin{equation}
\frac{C^{2}f(t)}{2m}+\frac{D^{2}m\omega_{0}^{2}g(t)}{2}+\frac{1}{2}\left( C%
\frac{\partial D}{\partial t}-D\frac{\partial C}{\partial t}\right) =\frac{%
m\omega^{2}}{2},  \label{17}
\end{equation}%
\begin{equation}
\frac{ACf(t)}{2m}+\frac{DBm\omega_{0}^{2}g(t)}{2}-\frac{1}{2}\left( D\frac{%
\partial A}{\partial t}-C\frac{\partial B}{\partial t}\right) =0.  \label{18}
\end{equation}
In principle, we can solve the coupled\textbf{\ }equations for deriving $%
A,B,C$ and $D$ when $f(t)$ and $g(t)$ are both given. As a result, the
Hamiltonian\textbf{\ }in Eq.(\ref{7}) can be turned into the
time-independent Hamiltonian of the standard harmonic oscillator.

As a concrete example, we derive the time-dependent Fresnel operator for the
CK Hamiltonian \cite{01,02}. The CK Hamiltonian is given by setting $%
f(t)=e^{-2\gamma t}$ and $g(t)=e^{2\gamma t}$ in Eq.(\ref{7})
\begin{equation}
H=e^{-2\gamma t}\frac{P^{2}}{2m}+e^{2\gamma t}\frac{1}{2}%
m\omega_{0}^{2}X^{2}.  \label{19}
\end{equation}
Correspondingly, these partial differential equations, from Eq.(\ref{15}) to
Eq.(\ref{17}), become%
\begin{equation}
\frac{A^{2}e^{-2\gamma t}}{2m}+\frac{B^{2}m\omega_{0}^{2}e^{2\gamma t}}{2}+%
\frac{1}{2}\left( A\frac{\partial B}{\partial t}-B\frac{\partial A}{\partial
t}\right) =\frac{1}{2m},  \label{20}
\end{equation}%
\begin{equation}
\frac{C^{2}e^{-2\gamma t}}{2m}+\frac{D^{2}m\omega_{0}^{2}e^{2\gamma t}}{2}+%
\frac{1}{2}\left( C\frac{\partial D}{\partial t}-D\frac{\partial C}{\partial
t}\right) =\frac{m\omega^{2}}{2},  \label{21}
\end{equation}%
\begin{equation}
-\frac{ACe^{-2\gamma t}}{2m}-\frac{m\omega_{0}^{2}e^{2\gamma t}DB}{2}+\frac {%
1}{2}\left( D\frac{\partial A}{\partial t}-C\frac{\partial B}{\partial t}%
\right) =0.  \label{22}
\end{equation}
From the point of view of dimensional analysis for Eq. (\ref{20}), we should
take $B=0,$ so $A=e^{\gamma t}.$ Due to $A\frac{\partial D}{\partial t}+D%
\frac{\partial A}{\partial t}-B\frac{\partial C}{\partial t}-C\frac{\partial
B}{\partial t}=0,$ we can further obtain $D=e^{-\gamma t}.$ Then from Eq.(%
\ref{22}) we know $C=m\gamma e^{\gamma t},$ namely
\begin{equation}
A=e^{\gamma t},\text{ }B=0,\text{ }C=m\gamma e^{\gamma t},\text{ }%
D=e^{-\gamma t}.  \label{22a}
\end{equation}
Substituting Eq.(\ref{22a}) into Eq.(\ref{21}) we obtain the frequency $%
\omega=\sqrt{\omega_{0}^{2}-\gamma^{2}}$. According to Eqs.\textbf{(}\ref{5}%
\textbf{) }and\textbf{\ (}\ref{22a}\textbf{)} the time-dependent Fresnel
operator for CK Hamiltonian takes the form%
\begin{equation}
F=\exp \left( \frac{\text{i}m\gamma}{2}X^{2}\right) \exp \left[ -\frac{\text{%
i}\gamma t}{2}(XP+PX)\right] ,  \label{23}
\end{equation}
which can convert the time-dependent CK Hamiltonian into the Hamiltonian of
the standard harmonic oscillator with a frequency $\omega$. Although this
kind of operators appeared in ref.\cite{05}, nevertheless, its physical
meaning as a particular Fresnel operator had not been noticed there, not to
mention how it was deduced.

\section{Wavefunction to the Schr\"{o}dinger equation of CK Hamiltonian}

Now from Eqs.(\ref{9}) and (\ref{14}) we know $\left \vert
\psi(t)\right
\rangle =F^{-1}\left \vert \phi \right \rangle .$ If $\left
\vert \phi \right \rangle =\left \vert n\right \rangle $, a number state in
the Fock space, is the eigenstate of $\mathcal{\hat{H}=}\frac{P^{2}}{2m}+%
\frac{1}{2}m\omega^{2}X^{2},$ the solution to the Schr\"{o}dinger equation
of CK Hamiltonian is%
\begin{equation}
\left \vert \psi(t)\right \rangle =\exp \left[ \frac{\text{i}\gamma t}{2}%
(XP+PX)\right] \exp \left( \frac{-\text{i}m\gamma}{2}X^{2}\right) \left
\vert n\right \rangle  \label{24}
\end{equation}
In the $\left \langle x\right \vert $ representation, $\left \langle
x\right
\vert X=x\left \langle x\right \vert ,$ using%
\begin{equation}
\left \langle x\right \vert \exp \left[ \frac{\text{i}\gamma t}{2}(XP+PX)%
\right] =e^{\gamma t/2}\left \langle e^{\gamma t}x\right \vert  \label{25}
\end{equation}
we know the wavefunction of $\left \vert \psi(t)\right \rangle $ is
\begin{align}
\left \langle x\right. \left \vert \psi(t)\right \rangle & =\left \langle
x\right \vert \exp \left[ \frac{\text{i}\gamma t}{2}(XP+PX)\right] \exp
\left( -\frac{\text{i}m\gamma}{2}X^{2}\right) \left \vert n\right \rangle
\notag \\
& =e^{\gamma t/2}\exp \left( -\frac{\text{i}m\gamma}{2}e^{2\gamma
t}x^{2}\right) \left \langle e^{\gamma t}x\right \vert \left. n\right \rangle
\label{26}
\end{align}
where
\begin{equation}
\left \langle e^{\gamma t}x\right \vert \left. n\right \rangle =\left(
\frac {1}{2^{n}n!}\sqrt{\frac{m\omega}{\pi \hbar}}\right) ^{1/2}e^{-\frac{%
m\omega e^{2\gamma t}}{2\hbar}x^{2}}H_{n}\left( \sqrt{\frac{m\omega}{\hbar}}%
e^{\gamma t}x\right) ,  \label{27}
\end{equation}
$H_{n}\left( x\right) $ is the single-variable Hermite polynomial and $\hbar$
is recovered. Clearly, the exact time-dependent wavefunction $\left \langle
x\right. \left \vert \psi(t)\right \rangle $ represents a squeezed number
state for the CK Hamiltonian model. As one can see from the above discussion
our method can be suitable for solving Schr\"{o}dinger equation of other
time-dependent oscillators.

Another advantage of our method is that the Wigner function of $\left\vert
\psi (t)\right\rangle $ can be concisely derived by the above time-dependent
Fresnel operator. The Wigner operator is defined by \cite{19}%
\begin{equation}
\Delta \left( x,p\right) =\int_{-\infty }^{+\infty }\frac{du}{2\pi }%
e^{iup}\left\vert x+\frac{u}{2}\right\rangle \left\langle x-\frac{u}{2}%
\right\vert .  \label{28}
\end{equation}%
its normal product form is \cite{11}%
\begin{align}
\Delta \left( x,p\right) & =\frac{1}{\pi }\colon
e^{-(x-X)^{2}-(p-P)^{2}}\colon   \notag \\
& =\frac{1}{\pi }\colon e^{-2(\alpha ^{\ast }-a^{\dagger })(\alpha
-a)}\colon   \notag \\
& \equiv \Delta \left( \alpha ,\alpha ^{\ast }\right) ,  \label{29}
\end{align}%
where $\alpha =\frac{1}{\sqrt{2}}\left( \sqrt{\frac{m\omega }{\hbar }}x+%
\text{i}\frac{p}{\sqrt{m\hbar \omega }}\right) .$ The Weyl ordered form $%
\Delta \left( x,p\right) $ is \cite{12}%
\begin{equation}
\Delta \left( x,p\right) =%
\begin{array}{c}
: \\
:%
\end{array}%
\delta (x-X)\delta (p-P)%
\begin{array}{c}
: \\
:%
\end{array}%
,  \label{30}
\end{equation}%
where $%
\begin{array}{c}
: \\
:%
\end{array}%
\begin{array}{c}
: \\
:%
\end{array}%
$ denotes operators' Weyl ordering. Noticing that the Weyl ordering has a
remarkable property, i.e., the order-invariance of Weyl ordered operators
under similar transformations \cite{12}, which means $F%
\begin{array}{c}
: \\
:%
\end{array}%
(\circ \circ \circ )%
\begin{array}{c}
: \\
:%
\end{array}%
F^{-1}=%
\begin{array}{c}
: \\
:%
\end{array}%
F(\circ \circ \circ )F^{-1}%
\begin{array}{c}
: \\
:%
\end{array}%
$as if the "fence" $%
\begin{array}{c}
: \\
:%
\end{array}%
\begin{array}{c}
: \\
:%
\end{array}%
$ did not exist. Then using Eqs.(\ref{6}), (\ref{22}) and (\ref{30}), the
Wigner function of the squeezed number state for CK Hamiltonian is%
\begin{eqnarray}
&&W_{n}(x,p)  \notag \\
&=&Tr\{\left\vert \psi (t)\right\rangle \left\langle \psi (t)\right\vert
\Delta \left( x,p\right) \}  \notag \\
&=&\left\langle n\right\vert F%
\begin{array}{c}
: \\
:%
\end{array}%
\delta (x-X)\delta (p-P)%
\begin{array}{c}
: \\
:%
\end{array}%
F^{-1}\left\vert n\right\rangle   \notag \\
&=&\left\langle n\right\vert
\begin{array}{c}
: \\
:%
\end{array}%
\delta \left( x-e^{-\gamma t}X\right) \delta \left[ p-e^{\gamma t}\left(
P-m\gamma X\right) \right]
\begin{array}{c}
: \\
:%
\end{array}%
\left\vert n\right\rangle   \notag \\
&=&\left\langle n\right\vert \Delta \left( x^{\prime },p^{\prime }\right)
\left\vert n\right\rangle   \notag \\
&=&\frac{e^{-2\left\vert \alpha ^{\prime }\right\vert ^{2}}}{\pi }%
(-1)^{n}L_{n}(4\left\vert \alpha ^{\prime }\right\vert ^{2}),  \label{31}
\end{eqnarray}%
where
\begin{equation}
x^{\prime }\equiv e^{\gamma t}x,\text{ }p^{\prime }\equiv m\gamma e^{\gamma
t}x+e^{-\gamma t}p,  \label{32a}
\end{equation}%
and
\begin{equation}
\alpha ^{\prime }\equiv \frac{1}{\sqrt{2}}\left( \sqrt{\frac{m\omega }{\hbar
}}x^{\prime }+\text{i}\frac{p^{\prime }}{\sqrt{m\hbar \omega }}\right) ,
\label{32b}
\end{equation}%
$L_{n}$ is the Laguerre polynomials. Especially, when $n=0,$ Eq.(\ref{31})
reduces to the Wigner function of the squeezed vacuum state%
\begin{equation}
W_{n=0}(x,p)=\frac{1}{\pi }\exp \left[ -\frac{\left( p+m\gamma e^{2\gamma
t}x\right) ^{2}}{m\hbar \omega e^{2\gamma t}}-\frac{m\omega e^{2\gamma t}}{%
\hbar }x^{2}\right] ,  \label{37}
\end{equation}%
For $n=1,$the Wigner function is%
\begin{align}
W_{n=1}(x,p)& =\frac{1}{\pi }\left[ \frac{2\left( p+m\gamma e^{2\gamma
t}x\right) ^{2}}{m\hbar \omega e^{2\gamma t}}+\frac{2m\omega e^{2\gamma t}}{%
\hbar }x^{2}-1\right]   \notag \\
& \times \exp \left[ -\frac{\left( p+m\gamma e^{2\gamma t}x\right) ^{2}}{%
m\hbar \omega e^{2\gamma t}}-\frac{m\omega e^{2\gamma t}}{\hbar }x^{2}\right]
\label{38}
\end{align}

Wigner functions expressed by Eq.(\ref{31}) are depicted in $(x,p)$ phase
space for different values of $t$ and $n$ when $\gamma=0.5$. Figs.1 and 2
respectively exhibit the cases of $n=0$ and $\ n=1$ for different time $t.$
Fig.1(a) shows the Wigner function of the vacuum state. Figs.1(b) and 1(c)
show that when time goes on, the form of Wigner functions in Gaussian
quickly becomes narrow in position space, but spreads widely in momentum
space, which implies squeezing mechanism involved in the CK Hamiltonian
model. In Fig.2 there is a negative region, which indicates the
nonclassicality of the squeezed state when $n\neq0$.
\begin{figure}[ptb]
\label{Fig1} \centering \includegraphics[width=8cm]{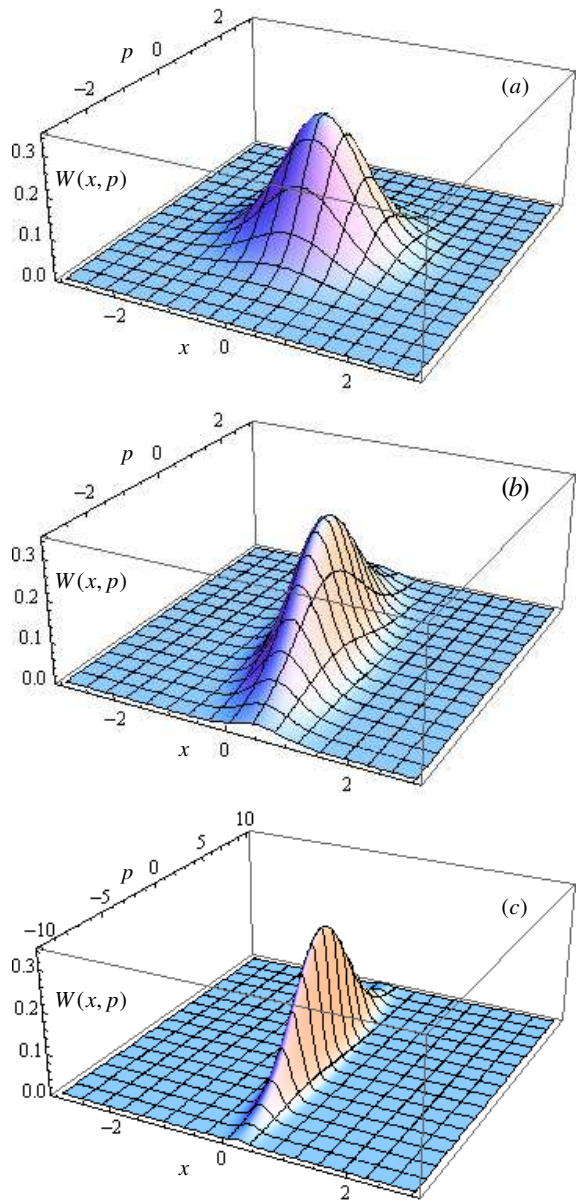}
\caption{(Color online) Wigner function of the squeezed state for the CK
Hamiltonian model for $n=0$ (a) $\protect\gamma=0,$\ $t=0;$ (b) $\protect%
\gamma=0.5,$\ $t=1;$(c)$\protect\gamma =0.5,$\ $t=3.$}
\end{figure}
\begin{figure}[ptb]
\label{Fig2} \centering \includegraphics[width=8cm]{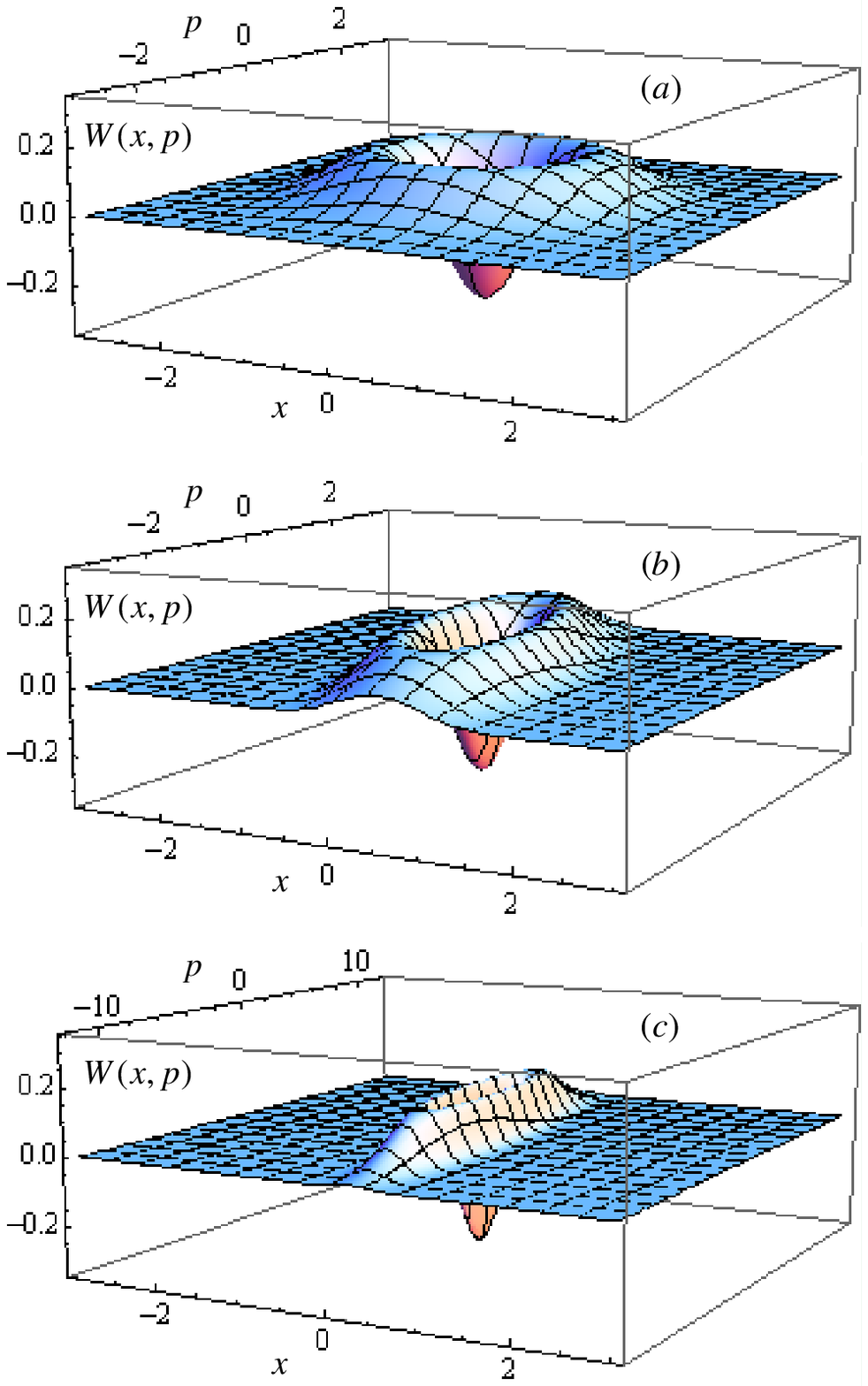}
\caption{(Color online) Wigner function of the squeezed state for the CK
Hamiltonian model for $n=1$ (a) $\protect\gamma=0,$\ $t=0;$(b)$\protect\gamma%
=0.5,$\ $t=1;$(c)$\protect\gamma=0.5,$\ $t=3.$}
\end{figure}

\section{Conclusion}

In summary, we have introduced the time-dependent Fresnel operator for
converting Caldirola-Kanai Hamiltonian into time-independent harmonic
oscillator Hamiltonian, the $A,B,C,D$ parameters involved in the Fresnel
operator are the solution to a set of the patrial differential equations set
up in the above mentioned converting process. In this way the dynamics of
Caldirola-Kanai Hamiltonian is solved, Our method may be suitable for
solving the Schr\"{o}dinger equation of other time-dependent oscillators.

\textbf{Acknowledgments }This\textbf{\ }work was supported by the the
National Natural Science Foundation of China under Grant Nos.10775097,
10874174 and Shandong Provincial Natural Science Foundation, China Grant
No.ZR2010AQ024.

\end{document}